\documentclass[reprint,amsmath,amssymb,aps,prx,longbibliography]{revtex4-2}

\usepackage[utf8]{inputenc} 
\usepackage[T1]{fontenc}
\usepackage[english]{babel} 
\usepackage[autostyle=true]{csquotes}
\usepackage[colorlinks=true, urlcolor=blue, linkcolor=blue,citecolor=blue]{hyperref}
\usepackage[dvipsnames]{xcolor} 
\usepackage{siunitx} 
\usepackage{braket} 
\usepackage{graphicx}
\usepackage{bm}
\usepackage{todonotes}
\usepackage{physics}
\usepackage{txfonts}
\usepackage{makecell}  

\definecolor{red1}{RGB}{175,0,0}
\definecolor{violet1}{RGB}{150,0,75}
\definecolor{green1}{RGB}{0,150,0}
\definecolor{blue1}{HTML}{1A237E}

\DeclareMathOperator{\e}{e}
\newcommand{\mi}{\mathrm{i}}

\newcommand{\bs}[1]{\boldsymbol{#1}}

\begin{document}
    
\title{Breathing and Fission of Magnetic Multi-solitons}

\author{G. Brochier,$^{1}$ Y. Li,$^{1}$ S. Wattellier,$^{1}$ S. Philips,$^{1}$ F. Rabec,$^{1}$  S. Nascimbene,$^{1}$ J. Dalibard,$^{1}$ J. Beugnon$^{1,2}$}

\email{beugnon@lkb.ens.fr}

\affiliation{$^{1}$Laboratoire Kastler Brossel,  Coll\`ege de France, CNRS, ENS-PSL
University, Sorbonne Universit\'e, 11 Place Marcelin Berthelot, 75005 Paris,
France}

\affiliation{$^{2}$Institut Universitaire de France (IUF)}

\date{\today}
\begin{abstract}
We report the deterministic experimental realization and controlled fission of magnetic multi-soliton states in a uniform quasi-one-dimensional immiscible two-component Bose gas. We explore the Manakov regime, where the spin dynamics is well described by the easy-axis Landau-Lifshitz equation (LLE). The gauge equivalence between the easy-axis LLE and the attractive nonlinear Schrödinger equation (NLSE) enables the direct construction of magnetic multi-solitons from the well-known NLSE solutions. We observe the two- and three- soliton states, which exhibit robust breathing in quantitative agreement with integrable theory. By introducing a weak, localized perturbation, we controllably break integrability and induce the splitting of a two-soliton into its fundamental constituents. This process reveals the composite structure of multi-soliton states and realizes an experimental analog of the inverse scattering transform.

\end{abstract}

\maketitle

\section{Introduction}

Solitons have been investigated in a broad variety of physical systems, reflecting the universality of nonlinear wave phenomena. Experimental realizations span classical hydrodynamics\,\cite{Maxworthy1976, Lake1977, Koop1981, Chabchoub2011}, nonlinear optics\,\cite{Hasegawa1973, Mollenauer1980, Kibler2010}, ultracold atomic gases\,\cite{Burger1999, Khaykovich2002, Strecker2002, Cornish2006, Becker2008, Mitchell2021}, magnetic media\,\cite{Mikeska1991}, Josephson junction systems\,\cite{Cuevas2014, De_santis2025}, and polymer chains\,\cite{Dauxois2004}. Across these diverse platforms, solitons manifest as robust, localized wavepackets governed by a balance between dispersion and nonlinearity.
In the 1960s, the development of powerful mathematical tools, such as the inverse scattering transform (IST), significantly broadened the notion of solitons in integrable systems. Within this framework, solitons appear as \enquote{modes} of the associated nonlinear partial differential equations (PDEs)\,\cite{Gardner1967, Lax1968, Zakharov1972_russ, Ablowitz1974}. This approach naturally encompasses time- or space-periodic solutions, including breathers\,\cite{Ma1979, Akhmediev1986} and prototypes of rogue waves\,\cite{Kibler2010, Akhmediev2016, Romero2024}, extending the family of soliton-like structures. 

A key property shared by all these solutions, and associated to the underlying integrability of the corresponding nonlinear equations, is their robustness under collisions, during which each soliton preserves its amplitude and velocity\,\cite{Zabusky1965, Frisquet2013, Nguyen2014}.
An intriguing situation arises when multiple overlapping solitons propagate at the same velocity, forming so-called multi-soliton states, which can be viewed as nonlinear superpositions of individual solitons. For instance, the sine-Gordon breather\,\cite{Cuevas2014, De_santis2024} corresponds to a bound soliton-antisoliton pair that oscillates in time while remaining spatially localized. More generally, integrable PDEs admit multi-soliton solutions in which several solitons coexist and interact, yet retain their individual identities.
In the case of the attractive nonlinear Schrödinger equation (NLSE), multi-soliton states can form without any associated binding energy between their constituent solitons. A celebrated example of $n$-soliton solution takes the simple form $u_n(x, t=0)=n u_1(x)$, where $u_1$ is the single soliton solution, and $n = 2, 3, \dots$\,\cite{Satsuma1974}. Such breather states have been observed experimentally in different platforms\,\cite{Chabchoub2013, Stolen1983, Luo2020}. 

This phenomenology is not limited to scalar wave equations but also extends to vectorial nonlinear models\,\cite{Bersano2018, Lannig2020} including the integrable Landau-Lifshitz equation (LLE). The LLE governs the spatiotemporal evolution of the magnetization in one-dimensional ferromagnetic spin chains in the continuous limit\,\cite{Kosevich1998}. This equation is widely used in condensed matter physics, for instance in the field of spintronics \cite{Lakshmanan11,Lenk11}. In the absence of dissipation, the LLE is integrable\,\cite{Takhtajan1977, Mikhailov1982} and exhibits deep connections with paradigmatic nonlinear equations such as the nonlinear Schrödinger and sine-Gordon equations\,\cite{Kosevich1998,Mikeska1991}. Its solitonic solutions are commonly referred to as magnetic solitons.

The experimental realization of such magnetic solitons has recently been proposed in the context of ultracold atomic gases\,\cite{Qu2016, Chai2022, Bresolin2023}. In particular, it was shown that the dynamics of two-component one-dimensional mixtures, governed by coupled NLSEs, can be mapped onto the LLE in the Manakov regime, when the inter- and intra-species interaction parameters are nearly equal. Magnetic solitons have since been observed in both miscible mixtures\,\cite{Farolfi2020, Chai2020} and immiscible ones\,\cite{Rabec2025}, corresponding to the easy-plane and easy-axis LLE cases, respectively. By contrast, multi-soliton solutions of the LLE remain far less explored than their NLSE counterparts, and their experimental observation is still lacking. Moreover, owing to the absence of binding energy, multi-soliton states are highly sensitive to weak integrability-breaking perturbations, which can induce their fission and thereby reveal their composite nature, as first evidenced in optical fibers\,\cite{Tai1988, Mucci2025}. The ability to engineer arbitrary external potentials in cold-atom platforms provides a promising route to control such splitting processes and thus to establish an experimental counterpart to the IST\,\cite{Marchukov2019}.

Creating and controlling the dynamics of multi-solitons in cold-atom systems thus constitute a crucial milestone in establishing this platform as a state-of-the-art experimental benchmark for exploring contemporary nonlinear physics. It opens the way to a unified experimental investigation of the Landau-Lifshitz, the nonlinear Schrödinger, and the sine-Gordon equations, thereby revealing the mathematical transformations that intertwine these paradigmatic nonlinear models.

In this article, we investigate multi-soliton states in a two-component immiscible quasi-one-dimensional Bose gas. This system is governed by the easy-axis Landau-Lifshitz equation (LLE), which supports magnetic soliton solutions. We explicit the gauge equivalence between the easy-axis LLE and the attractive NLSE, thereby enabling the construction of magnetic multi-soliton states directly from well-known NLSE soliton solutions. We experimentally realize these magnetic multi-solitons and characterize their breathing dynamics. By weakly and controllably breaking integrability, we also reveal the composite nature of two-soliton states and induce their deterministic fission. We directly map their solitonic content onto experimentally observable wavepackets, thus implementing an analog IST decomposition.

\section{IST and multi-solitons}

In this section, we briefly introduce the main features of the IST for the NLSE, which will allow us to interpret the studied multi-soliton states. The IST enables the systematic construction of solutions to certain nonlinear PDEs. It was originally developed in the context of the Korteweg-de Vries equation\,\cite{Gardner1967} and was subsequently extended to many other integrable systems\,\cite{Lax1968, Zakharov1972_russ, Ablowitz1974}, including the one-dimensional attractive (or focusing) NLSE. In dimensionless form, it reads:
\begin{equation}
    \mi u_t + u_{xx} + 2 \abs{u}^2 u = 0\,,
    \label{eq:NLSE}
\end{equation}
where subscripts denote partial derivatives with respect to the corresponding variables, and $u$ is a complex-valued field. The central idea of the IST is to recast the nonlinear PDE as a linear spectral problem whose solutions evolve trivially in time. 
In this sense, the IST plays a role analogous to the Fourier transform for linear PDEs. The integrability of Eq.~\eqref{eq:NLSE} can be formally expressed through the existence of a so-called Lax pair $(\hat{L}, \hat{A})$, defined by the auxiliary linear problems\,\cite{Ablowitz1974}:
\begin{subequations}
    \begin{alignat}{2}
        \Phi_x &= \hat{L}\, \Phi\,, \quad \hat{L}(\lambda) &&\equiv \mqty(-\mi \lambda & u \\ -u^* & \mi \lambda) \label{eq:IST_eigenL}\,,\\
        \Phi_t &= \hat{A}\, \Phi\,,  \quad \hat{A}(\lambda) &&\equiv  \mqty(-2\mi \lambda^2 + \mi \abs{u}^2 &\mi u_x + 2 \lambda u \\ \mi u_x^* - 2 \lambda u^* & 2\mi \lambda^2 - \mi \abs{u}^2)\, , \label{eq:IST_eigenA}
    \end{alignat}%
    \label{eq:IST_eigen}
\end{subequations}\unskip\ignorespaces
where $\lambda$ is \emph{a priori} complex-valued, $\cdot^*$ denotes complex conjugation, and the so-called Jost solutions $\Phi = \mqty(\phi_a & \phi_b)^\intercal$. The equivalence between this linear system and Eq.~\eqref{eq:NLSE} follows from the compatibility condition $\Phi_{xt} = \Phi_{tx}$.  Eq.~\eqref{eq:IST_eigenL} depends only  implicitly on time through the presence of the field $u$. We restrict our analysis to solutions of Eq.~\eqref{eq:NLSE} satisfying the vanishing boundary conditions $\lim_{x\to \pm\infty} |u(x,t)| = 0$. 

Eq.~\eqref{eq:IST_eigenL} can also be written as a non-Hermitian eigenvalue problem, with $\lambda$ the spectral parameter, and the Jost solutions being the eigenfunctions\,\cite{Zakharov1972_russ}. Real eigenvalues $\lambda$ correspond to scattering states, behaving as $\e^{\pm \mi \lambda x}$ at infinity. Complex eigenvalues can be viewed as bound states of the scattering problem, Eq.~\eqref{eq:IST_eigenL}, in the \enquote{potential} $u$, and each eigenvalue $\lambda_j$ is associated to a soliton. These solitons form a nonlinear basis of solutions. Importantly, the eigenvalues are conserved during the time propagation. For example, the well-known bright soliton solution of the NLSE:
\begin{equation}
    u(x,t) = \frac{\kappa \e^{\mi v x/2}}{\cosh[\kappa (x-vt)]} \e^{\mi \, (\kappa^2-v^2/4)\, t}\,,
    \label{eq:IST_bright_sol}
\end{equation}
corresponds to a single discrete eigenvalue $\lambda_1 = -v/4 + \mi \kappa/2$. The real part of $\lambda_j$ is directly related to the soliton velocity, while its imaginary part determines its amplitude and width. This relation provides a transparent explanation of the remarkable stability of solitons, in particular during collisions, since the eigenvalues satisfy $\partial_t \lambda_j = 0$. Owing to its integrability, the NLSE features an infinite set of conserved quantities, the first of which read for the bright soliton Eq.~\eqref{eq:IST_bright_sol}: $N_\text{at} = 2 \kappa$, $E = - 2\kappa^3/3$, the atom number (or mass) and energy respectively. In general, the field $u(x,t)$ can be reconstructed from the complete set of so-called scattering data, linked to the expression of the Jost solutions $\Phi$ at infinity. These data include notably the eigenvalues $\lambda_j$, but also encode the position and phases of the solitons. This reconstruction can be performed at any time $t$, using a Gelfand-Levitan-Marchenko integral equation (see Appendix~\ref{app:IST}).

In the following, we focus on a particular class of solutions characterized by $\lambda_j \in \mi \mathbb{R}$, corresponding to solitons at zero velocity. When the spatial profiles of individual solitons overlap, the phase factors associated to their scattering data, $\exp(4\mi \lambda_j^2 t)$, lead to a breathing dynamics, thanks to the linear character of the spectral problem (see Appendix~\ref{app:IST}). We refer to such solutions as multi-solitons.

An instructive class of multi-soliton states, on which we focus in this article, was introduced by Satsuma and Yajima\,\cite{Satsuma1974}, who showed that choosing an initial field equal to an integer multiple $n$ of a single bright soliton at rest leads to:
\begin{equation}
	u(x,0) = \frac{n \kappa}{\cosh(\kappa x)} ~,~ \lambda_j = \left\{\mi \frac{(2j-1) \kappa}{2} \right\}_{j \in \{1, \dots, n \} }\,,
	\label{eq:N_sol_NLSE}
\end{equation}
with no radiative component. A slight deviation of $n$ from an integer value leads to the appearance of a small radiative component. In this case, some atoms are transferred into dispersive waves, associated with $\lambda  \in \mathbb{R}$, and spread away from the breather during the evolution. For $n\in \mathbb{N}^*$, these multi-solitonic states verify $N_\text{tot} = \sum_j N_{\mathrm {at},j}$ and $E_\text{tot} = \sum_j E_j$, where $N_{\mathrm {at},j}$ and  $E_j$ denotes the mass and energy associated with a bright soliton of eigenvalue $\lambda_j$, respectively. Remarkably, these composite objects do not possess any binding energy. Because the eigenvalues $\lambda_j$ are commensurate, the resulting breathing dynamics is periodic. The case of the two-soliton ($n=2$) is associated with two eigenvalues, $\lambda_1 = \mi \kappa/2$ and $\lambda_2 = 3 \mi \kappa/2$. A single characteristic beating frequency thus appears in its breathing dynamics:
\begin{equation}
    f = \frac{1}{2\pi} \left(4\lambda_2^2 - 4\lambda_1^2 \right) = \frac{4}{\pi}\, \kappa^2 \,.
    \label{eq:sol_freq}
\end{equation}
This construction straightforwardly extends to larger values of $n$, yielding several commensurate frequencies and an overall periodic dynamics \footnote{Choosing irrational eigenvalues and $n\ge 3$ may on the contrary lead to a quasiperiodic behavior.}.

\section{Magnetic soliton and LLE}

\begin{figure}[t!!!]
    \begin{center}
	   \includegraphics[]{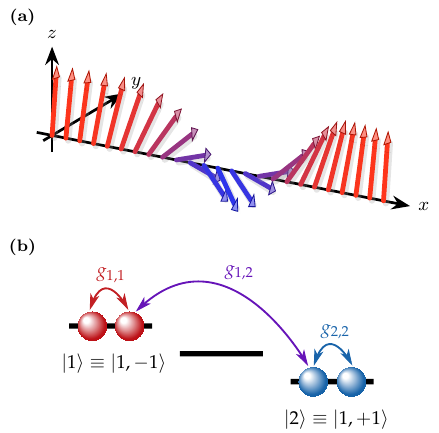} 
    \end{center}
\caption{\textbf{Analogy between an easy-axis ferromagnetic spin chain and a two-component Bose gas.} 
	\textbf{(a)} Schematic representation of a one-dimensional easy-axis ferromagnetic spin chain, governed by the LLE in the continuous limit. The magnetization has a constant modulus. It points along $+\bs{e}^{(z)}$ at infinity, while it bends toward the opposite direction in the middle, representing a magnetic soliton.
	\textbf{(b)} Experimental realization of an effective spin degree of freedom using two hyperfine states of the $F=1$ ground manifold of $^{87}$Rb. The system is characterized by intra- and inter-species interaction constants $g_{i,j}$. Owing to the states symmetry, $g_{1,1} = g_{2,2} \equiv g$. This immiscible two-component Bose gas maps onto an easy-axis ferromagnet, with magnetic solitons encoded in the local spin imbalance and relative phase between the components.}
    \label{fig:Scheme}
\end{figure}

We are interested in this work in the relation between the NLSE and another integrable one-dimensional nonlinear equation, the Landau-Lifshitz equation (LLE), which can be implemented in cold atomic platforms. We focus on the so-called easy-axis anisotropy case:
\begin{equation}
    \bs{M}_t = \left[\bs{M}_{xx} + \left(\bs{M} \cdot \bs{e}^{(z)} \right)\, \bs{e}^{(z)} \right]\times \bs{M}\,,
    \label{eq:LLE}
\end{equation}
where $\bs{e}^{(z)}$ is the unit vector along the $z$ axis. This equation describes the time evolution of the magnetization $\bs{M}(x,t)$ in a ferromagnetic spin chain with a preferred axis along which the magnetization aligns in the ground state, as sketched in Fig.~\ref{fig:Scheme}(a). In the following, we restrict ourselves to the uniform case, $\norm{\bs{M}} = 1$, which naturally leads to a parametrization of the magnetization in spherical coordinates, $\bs{M} = \mqty(\sin(\theta) \cos(\varphi) & \sin(\theta) \sin(\varphi) & \cos(\theta))^\intercal$. Eq.~\eqref{eq:LLE} then admits solitonic solutions, with a characteristic width $\kappa$ and a velocity $v$, called magnetic solitons (see Ref.~\cite{Kosevich1990} for a comprehensive review, and Appendix~\ref{app:Mag_soliton}).

The existence of such solutions suggests the integrability of the LLE. This property was indeed established using the IST for general anisotropies\,\cite{Mikhailov1982,Rodin1984}, and explicitly formulated for the easy-axis case in Ref.~\cite{Borovik1988}. A Lax pair associated with Eq.~\eqref{eq:LLE} can be written as:
\begin{subequations}
    \begin{align}
        \hat{L} &= \frac{1}{4} \left[\hat{\sigma}^{(z)}, \hat{M} \right] - \mi \lambda \hat{M} \,,\\
        \hat{A} &= \frac{1}{4} \left[\hat{\sigma}^{(z)}, 2\lambda \hat{M}+\mi\hat{M} \hat{M}_x \right] - \frac{\mi}{2} M^{(z)} \hat{\sigma}^{(z)} + \lambda \hat{M} \hat{M}_x - 2 \mi \lambda^2 \hat{M}\,,
    \end{align}
    \label{eq:Lax_LLE}
\end{subequations}\unskip\ignorespaces
where $[\cdot,\cdot]$ is the commutator, $\hat{M} = \sum_i M^{(i)} \hat{\sigma}^{(i)}$, and $\hat{\sigma}^{(i)}$ denotes the Pauli matrices.

Beyond its integrability, the easy-axis LLE admits a gauge equivalence with the attractive NLSE, Eq.~\eqref{eq:NLSE}, which establishes a direct correspondence between magnetic solitons and bright soliton solutions of the NLSE\,\cite{Nakamura1982,Kundu1983}. More formally, the Lax pairs associated with these two nonlinear equations can be related through a gauge transformation defined by a unitary matrix $\hat{G}$, independent of the spectral parameter $\lambda$\,\cite{Zakharov1979}:
\begin{subequations}
    \begin{align}
       \hat{L} &= \hat{G} \hat{L}' \hat{G}^{-1} + \hat{G}_x \hat{G}^{-1} \,, \label{eq:gauge_eq_L}\\ 
       \hat{A} &= \hat{G} \hat{A}' \hat{G}^{-1} + \hat{G}_t \hat{G}^{-1} \,,\label{eq:gauge_eq_A}
    \end{align}
    \label{eq:gauge_eq}
\end{subequations}\unskip\ignorespaces
where primed quantities refer to the NLSE. Such a transformation induces a one-to-one correspondence between the Jost solutions of the two scattering problems:
\begin{equation}
    \left\{
    \begin{aligned}
        \Phi_x &= \hat{L} \Phi \\
        \Phi_t &= \hat{A} \Phi 
    \end{aligned}
    \right.
    \quad \Longleftrightarrow \quad
    \left\{
    \begin{aligned}
        (\hat{G}^{-1} \Phi)_x &= \hat{L}' (\hat{G}^{-1} \Phi) \,,\\
        (\hat{G}^{-1} \Phi)_t &= \hat{A}' (\hat{G}^{-1} \Phi) \,,
    \end{aligned}
    \right.
    \label{eq:jost_equiv}
\end{equation}
thereby relating their respective scattering data. More specifically, Eq.~\eqref{eq:gauge_eq} can be recast as a differential equation for the matrix $\hat{G}$\,\cite{Kotlyarov1981} (see Appendix \ref{app:Gauge} for details). In the particular case where the NLSE field satisfies $u(x) \in \mathbb{R}$, this relation simplifies to:
\begin{equation}
    u(x) = \frac{1}{2} \left(\theta_x + \sin{\theta}\right) \qq{and} \varphi = 0\,,
    \label{eq:diff_eq_theta}
\end{equation}
which will play a central role in the following. The requirement $u(x, t=0) \in \mathbb{R}$ is satisfied for the NLSE multi-solitons given by Eq.~\eqref{eq:N_sol_NLSE}. Then, Eq.~\eqref{eq:diff_eq_theta} provides a direct route to construct LLE multi-soliton solutions from the well-known NLSE ones, imposing the natural boundary conditions $\lim_{x\to\pm \infty} \hat{M} = \hat{\sigma}^{(z)}$, that is $\lim_{x\to\pm \infty} \theta = 0\,[2 \pi]$.

\begin{figure*}[t!!!]
    \begin{center}
	   \includegraphics[]{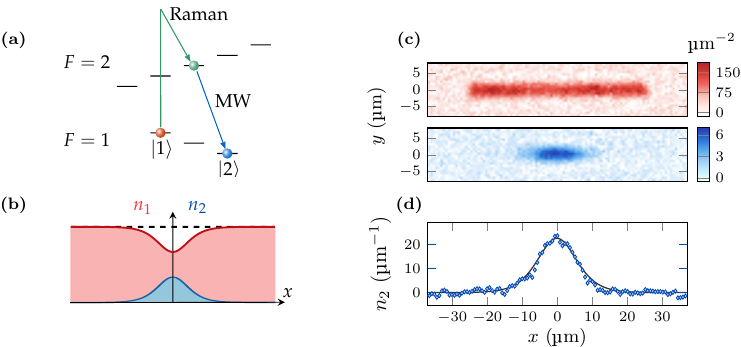} 
    \end{center}
\caption{\textbf{Realization of an arbitrary spin mixture.} 
	\textbf{(a)} Schematic of the relevant levels and transfer process. Two co-propagating Raman beams transfer part of the cloud to the intermediate $\ket{F=2, m_F=0}$ state in a spatially-resolved way. A global micro-wave $\pi$-pulse is then applied to drive the atoms to $\ket{2}$.
	\textbf{(b)} Schematic densities in each spin component at the end of the preparation sequence. Starting from a homogeneous gas, atoms are selectively transferred from $\ket{1}$ to $\ket{2}$, such that the total density remains uniform. 
	\textbf{(c)} Spin-selective absorption imaging of either component. Top: bath component $\ket{1}$. Bottom: minority component $\ket{2}$. 
	\textbf{(d)} One-dimensional density obtained by integrating the images along the $y$ direction. The solid line is a fit to the data using the profile of a magnetic soliton at rest, given by Eq.~\eqref{eq:mag_soliton}.}
    \label{fig:Exp1}
\end{figure*}

Two-component Bose mixtures constitute a suitable platform to realize and compare multi-solitons governed by both the NLSE\,\cite{Bakkali2023} and the LLE\,\cite{Rabec2025}. They are described by two fields $u_1$ and $u_2$, which satisfy coupled NLSEs:
\begin{subequations}
	\begin{align}
		&\mi (u_1)_t + (u_1)_{xx} - \frac{g_{1,1}}{g_s} \abs{u_1 }^2 u_1 - \frac{g_{1,2}}{g_s} \abs{u_2 }^2 u_1 = 0\,, \\
		&\mi (u_2)_t + (u_2)_{xx} - \frac{g_{2,2}}{g_s} \abs{u_2 }^2 u_2 - \frac{g_{1,2}}{g_s} \abs{u_1 }^2 u_2 = 0\,,
	\end{align}
	\label{eq:coupled_NLSE}
\end{subequations}\unskip\ignorespaces
where we have introduced $g_s = g_{1,2} - \sqrt{g_{1,1} g_{2,2}}$, and $g_{i,j}$ is the interaction constant between particles in state $i$ and $j$, illustrated in Fig.~\ref{fig:Scheme}(b). In this article, all equations are written in dimensionless forms. We provide their relation to the physical ones in Appendix~\ref{app:LLE}. 

In the Manakov regime, where all interaction constants are nearly equal, the total density can be approximated as constant, allowing us to parametrize the two-component fields as\,\cite{Qu2016, Congy2016}:
\begin{equation}
    \mqty(u_1 \\ u_2 ) = \e^{\mi \chi/2} \mqty(\cos(\theta/2)\e^{+\mi\varphi/2} \\ \sin(\theta/2)\e^{-\mi\varphi/2} )\,,
    \label{eq:psi_param}
\end{equation}

The global phase $\chi(x,t)$ can then be expressed as a function of the angular fields $\theta(x,t), \varphi(x,t)$, through Eq.~\eqref{eq:Phi_param}, and the resulting equation of motion for $\bs{M}$ coincides with Eq.~\eqref{eq:LLE}. Consequently, the local population imbalance and relative phase of the two-component fields directly encode the evolution of the magnetization in an effective spin chain. Two-component quantum gases are thus a pristine platform to implement the LLE.  In the regime of small spin excitations, corresponding to a low-depletion limit $n_2 \ll n_1(x\to\infty) = 1$, where $n_i = \abs{u_i}^2$, the two-component mixture is predominantly in  $\ket{1}$, with only a small population in $\ket{2}$. Then, we can write $u_2 = (M^{(x)} + \mi M^{(y)})/2$, which obeys, to leading order, a single-component attractive NLSE\,\cite{Kosevich1990}

In the experiments described below, we explore magnetic multi-solitons at rest associated to the combination of Eqs.~\eqref{eq:N_sol_NLSE} and \eqref{eq:diff_eq_theta}. They are fully defined by their order $n$ and their inverse width $\kappa$. We restrict ourselves to situations where $\kappa < \kappa_c(n)\equiv 1/(2n-1)$, for which the magnetic soliton wave functions have a uniform phase at $t=0$\,\footnote{The mass of the magnetic (multi-)soliton diverges at $\kappa = \kappa_c$, following Eq.~\eqref{eq:mag_soliton_N}. For $\kappa > \kappa_c$, one can still recourse to Eq.~\eqref{eq:diff_eq_theta} to construct solitonic solutions, but they exhibit phase jumps on $\varphi$.}. The NLSE limit is obtained for the low-depletion regime, corresponding to $\kappa \ll 1$.


%
\begin{figure*}[t!!!]
    \begin{center}
	   \includegraphics[]{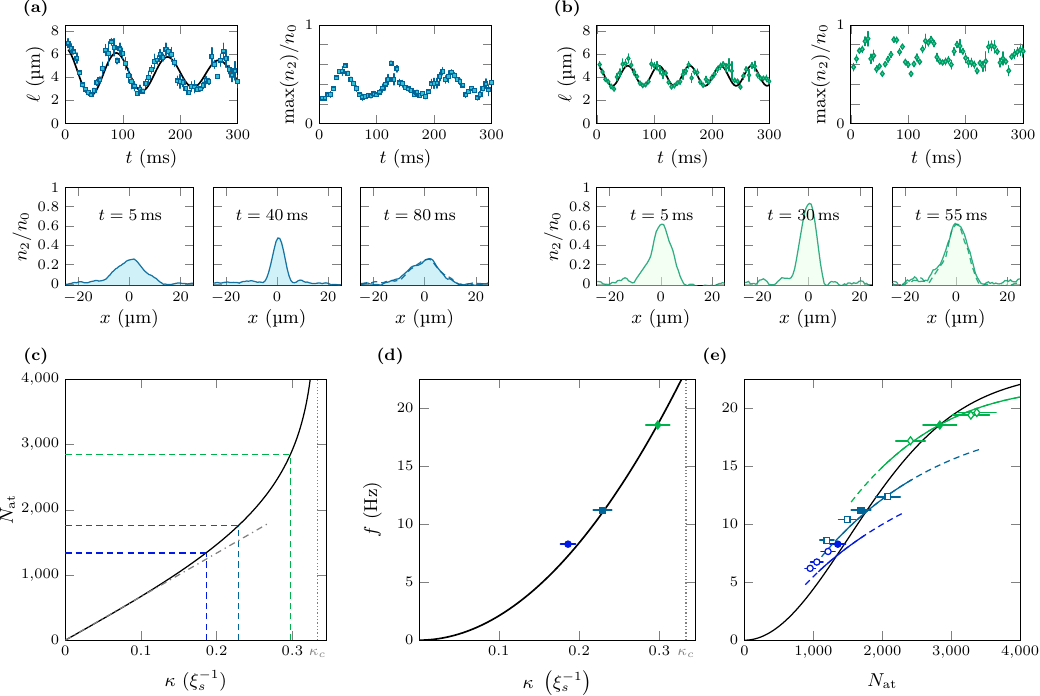} 
    \end{center}
\caption{\textbf{Breathing behavior of two-solitons.} 
	\textbf{(a)} Example of a breathing two-soliton for $\kappa=0.23(1)$. Shown are the time evolution of the soliton width (top left) and depletion (top right). Error bars correspond to the 1-$\sigma$ statistical uncertainty obtained from typically 15 repetitions of the experiment. The width $\ell(t)$ is obtained from a fit of the density at every time of the evolution with a function $\propto 1/\cosh^2(x/\ell)$. The solid line in the top panel corresponds to a damped sinusoidal fit, from which we extract the frequency. The decay of the oscillation amplitude is attributed to the presence of a small residual potential. The bottom panel shows the integrated atomic densities over the length $L$ of the segment, at times corresponding to $t \approx 0, T/2, T$, where $T$ denotes the breathing period. In the last panel, the initial profile (dashed line) is overlaid to emphasize the periodic breathing dynamics.
	\textbf{(b)} Same as in \textbf{(a)}, but for a two-soliton in the strongly depleted regime with $\kappa=0.30(1)$.
	\textbf{(c)} Evolution of the atom number in the two-soliton with $\kappa$, obtained by adding the atom numbers of the two constituent one-solitons given in \eqref{eq:mag_soliton_N}. The three dashed lines correspond to the three values $\kappa_i$ investigated experimentally and to  the associated  atom numbers. The soliton mass $N_{\mathrm at}$ diverges at $\kappa=\kappa_c(2)=1/3$. The dash-dotted line indicates the low-depletion NLSE limit.
	\textbf{(d)} Measured breathing frequency for two-solitons with different inverse widths $\kappa$. The solid black line indicates the frequency of the exact two-soliton as $\kappa$ is varied continuously. 
	\textbf{(e)} Breathing frequency in the vicinity of the exact two-soliton solution. For each  $\kappa_i$, the wavepacket amplitude is varied around its expected value within a small range. The colored lines represent the predicted frequencies with no adjustable parameters; they are dashed when the predicted quantity of radiated atoms is below 4\,\%, solid when it is below  1\,\%, and omitted otherwise.  The filled symbols correspond to the data points shown in \textbf{(d)}. For each investigated value $\kappa_i$, they are determined by selecting the point whose atom number $N_{\mathrm{at}}(\kappa_i)$  is closest to the predicted value indicated in \textbf{(c)} by the horizontal dashed lines.}
    \label{fig:2soliton}
\end{figure*}

\section{Experimental system}

Our experimental platform allows us to realize versatile two-component Bose mixtures in the Manakov regime. Here, we study one-dimensional Bose gases of $^{87}\mathrm{Rb}$ atoms of mass $m$, whose motion is confined by blue-detuned optical dipole traps formed by laser beams at a wavelength $\lambda = \SI{532}{\nano\meter}$. The atoms are loaded into a single dark fringe of a tunable optical lattice along the vertical direction $z$, producing an approximately harmonic confinement with a frequency of $\omega_z/2\pi = \SI{4.3\pm0.1}{\kilo\hertz}$\,\cite{Ville2017}. The resulting cloud is thus in a quasi-2D geometry\,\cite{Petrov2000}. The in-plane confinement is generated using a digital micro-mirror device (DMD). A flat-bottom pattern  is projected onto the atoms through a high-numerical-aperture objective with a resolution of approximately $\SI{1}{\micro\meter}$. This results in a box-like potential along $x$ with a length $L = \SI{50}{\micro\meter}$, and a smooth profile with a characteristic width of $\SI{3}{\micro\meter} = 2 \sigma_y$ along $y$. We obtain gases with a uniform linear density around $n_0 = \SI{550}{\per\micro\meter}$. The temperature of the sample is $\lesssim \SI{20}{\nano\kelvin}$, and we assume in the following that our experiments are well-described by a zero-temperature model.

A quantization magnetic field of approximately $\SI{1.2}{G}$ is applied along the $y$ axis, lifting the degeneracy of the Zeeman sublevels. The atoms are prepared in the hyperfine ground state $\ket{1} \equiv \ket{F=1, m_F=-1}$, where $F$ denotes the total angular momentum and $m_F$ its projection along the quantization axis. 
A two-step protocol, sketched in Fig.~\ref{fig:Exp1}(a), then transfers a fraction of the atoms to  $\ket{2} \equiv \ket{F=1, m_F=+1}$. It consists in a two-photon Raman pulse\,\cite{Zou2021, Bakkali2021} to the intermediate state $\ket{F=2, m_F=0}$, with the laser operating at a wavelength $\lambda = \SI{790}{\nano\meter}$, followed by a global microwave transfer. The Raman pulse has a spatially controlled intensity, thanks to another SLM. The result of this process is illustrated in Fig.~\ref{fig:Exp1}, where we sketch the two spin densities, and provide experimental images of both spin species, demonstrating the well-controlled imprinting of arbitrary spin textures. In practice, the waveform is computed numerically for each chosen value of $\kappa$ by combining Eq.~\eqref{eq:N_sol_NLSE} and Eq.~\eqref{eq:diff_eq_theta}, and imprinted on the atoms. After time evolution the density distribution is probed using spin-selective absorption imaging.

For the studied mixture, the intra- and interspecies interaction strengths are nearly equal, $g_{1,1} = g_{2,2} \equiv g \approx g_{1,2}$. We therefore obtain the spin interaction parameter $g_s = g_{1,2} - g$, which satisfies $|g_s| \ll g$. In our system, $g_s > 0$, implying that the mixture is weakly immiscible. This parameter defines a characteristic length scale, $\xi_s = \hbar / \sqrt{2 m g_s n_0} \approx \SI{1.5}{\micro\meter}$, known as the spin healing length, where $\hbar$ is the reduced Planck constant. Since $\xi_s$ is much larger than the one-component healing length, $\xi = \xi_s \sqrt{g_s/g}$, the density and spin dynamics are effectively decoupled, with the latter occurring on a lower energy scale. In addition, $\xi_s \approx \sigma_y$, allowing us to approximate the spin dynamics within a one-dimensional description.
Within this framework, we write $g_s = 2\hbar\omega_\perp \,(a_{1,2} - a)\, $, with $\omega_\perp = \sqrt{\omega_y \omega_z}$. Our calibration of $g_s$ gives $\omega_y / 2\pi = \SI{28 \pm 5}{\hertz}$\,\cite{Rabec2025}. Here, the scattering lengths are $a = (100.4 - 0.18)\, a_B$ and $a_{1,2} = (101.3 + 0.18)\, a_B$, where $a_B$ denotes the Bohr radius\,\cite{Van_kempen2002}. The small corrections $\pm 0.18\, a_B$ arise from magnetic dipole-dipole interactions in the quasi-2D geometry\,\cite{Zou2020a}.

\section{N-solitons: from NLSE to LLE}

We now report on the experimental observation of magnetic two- and three-soliton states in our atomic platform. We present the study of solitons at rest given by Eq.~\eqref{eq:N_sol_NLSE} in the low-depletion (NLSE) regime and their counterpart obtained from Eq.~\eqref{eq:diff_eq_theta} in the more general LLE framework. 

A key consequence of the gauge equivalence between the attractive NLSE and the easy-axis LLE is that the scattering eigenvalues $\lambda_j$ associated to solitons are preserved by the transformation. This result can be seen from Eq.~\eqref{eq:jost_equiv}. The time evolution of the scattering data follows simple laws, as in the NLSE case. The eigenvalues remain constant, and the associated scattering data acquire phase factors, which are given in the LLE case by $\exp[4 \mi (\lambda_j^2 + 1)t]$\,\cite{Borovik1988}, the global phase factor $\e^{4\mi t}$ being irrelevant in the following. Consequently, the characteristic breathing of magnetic two-soliton states is governed by the same formula, Eq.~\eqref{eq:sol_freq}, in both the NLSE and LLE frameworks. In physical units, the breathing frequency reads $f=2\hbar \kappa^2 / (\pi m)$. 

We study in Fig.~\ref{fig:2soliton} the breathing dynamics of two-solitons with eigenvalues $\lambda_1 = \mi \kappa/2$ and $\lambda_2 = 3\mi \kappa/2$, for different values of $\kappa$. We show the density profiles and the time evolution of their width and of the depletion max$(n_2)/n_0$ for two different values of $\kappa$ in Fig.~\ref{fig:2soliton}(a,b) and extract the corresponding breathing frequencies. 
We show in Fig.~\ref{fig:2soliton}(c) the predicted atom number in the exact two-soliton as a function of $\kappa$. For three chosen values of $\kappa$, we experimentally adjusted the atom number in the wavepacket  to this predicted value and determined the breathing frequencies reported in Fig.~\ref{fig:2soliton}(d). Excellent agreement is obtained with the prediction of  Eq.~\eqref{eq:sol_freq}, without any adjustable parameter. 

For a given value of the inverse width $\kappa$ of the wavepacket, small changes in the atom number only lead to a slightly shifted frequency of the width oscillations with no detectable radiation of atoms outside the wavepacket. We investigate this effect in Fig.~\ref{fig:2soliton}(e), where we plot the measured oscillation frequency as a function of the atom number in the wavepacket for the three studied values of $\kappa$. The experimental data  are in good agreement with the predicted frequency when varying $n$ around 2 (solid colored lines). Over the explored range of $N_\mathrm{at}$, less than 1\% of the atoms are expected to be radiated, which is below our experimental resolution.

\begin{figure}[t!!!]
    \begin{center}
	   \includegraphics[]{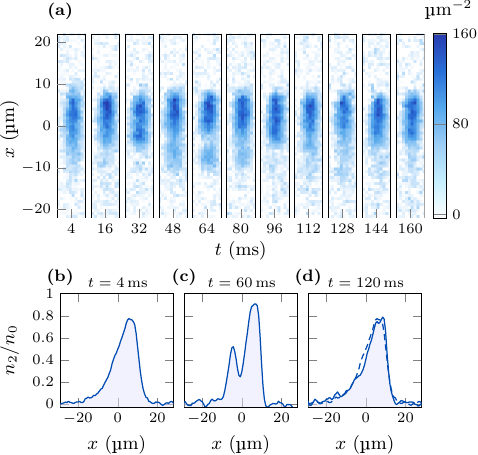} 
    \end{center}
\caption{\textbf{Observation of a magnetic three-soliton.} \textbf{(a)} Absorption images of a magnetic three-soliton, whose initial profile is obtained from Eq.~\eqref{eq:diff_eq_theta} using $u(x)=3\kappa/\cosh(\kappa x)$ with $\kappa=\num{0.19\pm0.01}$. \textbf{(b-d)} One-dimensional density profiles at different evolution times. A local minimum forms around half a period of the oscillation ($t \approx60\,$ms). A pronounced asymmetry is observed, originating from the relative spatial shifts of the constituent solitons induced by the gauge transformation. In \textbf{(d)}, the initial profile (dashed line) is overlaid to emphasize the periodic breathing dynamics.}
    \label{fig:3-soliton}
\end{figure}

The above discussion can be extended to higher-order multi-soliton states. Fig.~\ref{fig:3-soliton}(a) shows the time evolution of a three-soliton for $\kappa=0.19(1)$, close to $\kappa_c(3)=0.2$. The dynamics is periodic with a period of $\approx 120\,$ms. It is in good agreement with the expected value $\SI{133\pm14}{\milli\second}$, where the error follows from the uncertainty in the value of $\kappa$ for the imprinted wavepacket.  The presence of three discrete eigenvalues in the IST spectrum gives rise to three commensurate characteristic frequencies in the time evolution, at the origin of the observed more intricate dynamics.

In contrast to the NLSE three-soliton given by Eq.\,\eqref{eq:N_sol_NLSE}, the  wavepacket imprinted at time $t=0$, obtained thanks to the gauge mapping from NLSE to LLE,  displays a clear asymmetry. While not specific to the three-soliton case, this asymmetry becomes increasingly apparent in the large-depletion regime, \textit{i.e.} when $\kappa \to \kappa_c$. It originates from the fact that the gauge transformation, while preserving the discrete eigenvalues $\lambda_j$, does not preserve the remaining scattering data, encoded in the norming constants $c_j$ of the Jost functions $\Phi$. 
These coefficients determine the relative phases and spatial offsets of the individual solitons composing a multi-soliton state. Consequently, the gauge mapping induces nonlinear shifts in the constituent solitons position, which manifest as asymmetric density profiles. Symmetric magnetic multi-soliton states can also be constructed for the same set of eigenvalues $\lambda_j$, provided the norming constants $c_j$ of the initial NLSE solution $u$ are appropriately chosen. Achieving this requires using the general multi-soliton expression, Eq.~\eqref{eq:IST_multi_sol}, as the input of the differential relation Eq.~\eqref{eq:diff_eq_theta}, rather than the particular Satsuma-Yajima profile of Eq.~\eqref{eq:N_sol_NLSE}.

\section{Fission of a two-soliton}

Physical systems are rarely perfectly isolated, and their governing equations often include small additional terms representing external effects or imperfections. For integrable equations, such perturbations generally break exact integrability. Nevertheless, the IST framework remains a powerful tool (see, \textit{e.g.}, the review\,\cite{Kivshar1989}), as it allows one to follow the evolution of soliton parameters, and to understand how multi-soliton states respond to weak deviations from the ideal model.

Perturbations of the NLSE can be written as:
\begin{equation}
    \mi u_t + u_{xx} + 2\abs{u}^2u = \epsilon R[u] \qq{with \emph{e.g.}} \epsilon R[u] = V_\mathrm{ext}\, u\,,
    \label{eq:nlse_perturb}
\end{equation}
where $R$ may represent, for instance, particle loss or the effect of an external potential $V_\mathrm{ext}$, of order $\epsilon$. In this case, the scattering eigenvalues $\lambda_j$, which characterize the solitons, acquire a time dependence\,\cite{Karpman1977_russ}:
\begin{equation}
    \pdv{\lambda_j}{t} \propto \epsilon \int_{-\infty}^{+\infty} R^*[u] \phi_a^2 + R[u] \phi_b^2 \dd{x}\,,
    \label{eq:perturb_lambda}
\end{equation}
where the proportionality factor includes some coefficients related to the scattering data.
Unlike in the integrable IST framework, knowledge of the full Jost functions $\Phi(x,t;\lambda_j) = \mqty(\phi_a & \phi_b)^\intercal$ is required to compute the evolution of $\lambda_j$. While approximations may lead to analytical results for single-soliton states\,\cite{Karpman1977_russ} or for well-separated two-solitons\,\cite{Karpman1981}, numerical approaches are generally necessary to study arbitrary multi-soliton configurations\,\cite{Okamawari1995, Prilepsky2007}.

As multi-solitons do not possess any binding energy, any differential velocity acquired by one of the constituent solitons thus leads to a separation of the overall wavepacket. Conversely, the ability to impart controlled velocities ($\propto \mathrm{Re}(\lambda_j$)) to individual solitons without modifying their amplitudes ($\propto \mathrm{Im}(\lambda_j$)) effectively enables the realization of an analog IST, as it reveals the constituent solitons forming the initial multi-soliton state\,\cite{Tai1988, Sakaguchi2004}.

\begin{figure}[t!!!]
    \begin{center}
	   \includegraphics[]{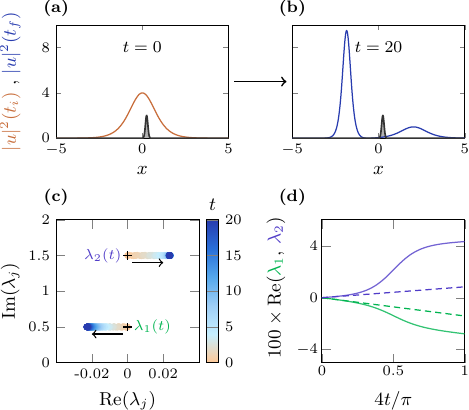} 
    \end{center}
    \caption{\textbf{Fission protocol of a two-soliton.}
    Numerical simulation of the time evolution of the eigenvalues for a two-soliton state of NLSE with initial wavepacket $u(x)=2/\cosh(x)$, in the presence of the localized perturbation $V_\mathrm{ext}(x)$. The parameters are $\sigma=0.1$, $\epsilon=0.02$, and $x_0=0.25$. 
    \textbf{(a-b)} Illustration of the initial and final densities, where the two solitons are separated. The external potential is shown as a filled black line and magnified by a factor $100$ for clarity. 
    \textbf{(c)}  Time evolution of the complex IST spectrum. The initial discrete eigenvalues, $\lambda_1=\mi/2$ and $\lambda_2=3\mi/2$, are marked with a cross, and their time-evolution under the action of the perturbation is color-coded. The imaginary parts remain approximately constant. 
    \textbf{(d)} Early-time evolution of the real parts of the two eigenvalues. Time is rescaled in units of the two-soliton breathing period. Solid lines correspond to the numerical prediction, solving Eq.\,\eqref{eq:IST_eigenL}. Dashed lines show the result of perturbation theory [Eq.~\eqref{eq:perturb_lambda} at $t=0$].}
    \label{fig:splitting_num}
\end{figure}
\begin{figure}[t!!!]
    \begin{center}
	   \includegraphics[width=8.6cm]{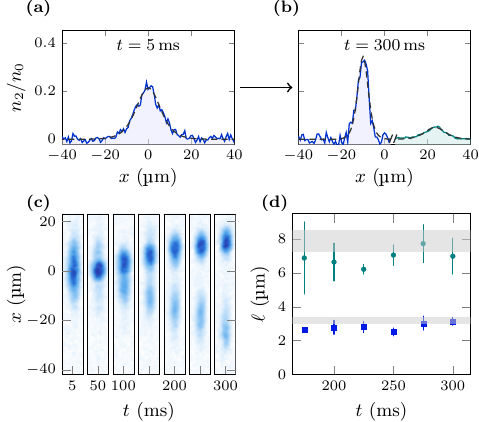} 
    \end{center}
    \caption{\textbf{Experimental fission of a two-soliton.}
    \textbf{(a-b)} Illustration of the initial and final densities, when the two solitons are separated. Due to their different amplitudes, distinct parameter sets are used to image the two wavepackets. They are displayed in different colors in  \textbf{(b)}.
    The black dashed lines correspond to $\propto 1/\cosh^2(x/\ell)$ fits.
    \textbf{(c)} Averaged absorption images illustrating the splitting of a two-soliton with $\kappa=0.20(1)$. We clearly observe the formation of two wavepackets, each one being a magnetic soliton.   
    \textbf{(d)} Fitted widths of the wavepackets, which remain approximately constant after the fission. The gray areas indicate the expected widths of magnetic solitons for the corresponding atom numbers and velocities. Their thickness indicates the uncertainty in our calibration of the interaction constant $g_s$. Error bars correspond to the 1-$\sigma$ statistical uncertainty obtained from typically 10 repetitions of the experiment.} 
    \label{fig:splitting_exp}
\end{figure}

Being able to controllably break the integrability allows one to induce the fission of multi-solitons. Following the proposal of Ref.~\cite{Marchukov2019}, we consider a localized potential $R[u]=f\,u$ with $f(x)= \exp[-{(x-x_0)^2}/{\sigma^2}]$ and $\sigma < 1/\kappa$, see Fig.~\ref{fig:splitting_num}(a-b). At the initial time, this potential induces a shift in the real part of the eigenvalues only, following Eq.~\eqref{eq:perturb_lambda}. For $\epsilon \ll 1$, their imaginary part, and thus their amplitudes, remain constant at all times, as illustrated by the numerical results shown in Fig.~\ref{fig:splitting_num}(c-d). The dynamics reverts to that of an integrable system once the solitons have propagated outside the region of the perturbation.


Our experimental fission protocol starts from the deterministic preparation of a magnetic two-soliton, as described in Fig.~\ref{fig:2soliton}. In order to have a longer available length to study the solitons motion, the geometry of the bath is changed from a $\SI{50}{\micro\metre}$-long segment  to a ring of radius $\SI{20}{\micro\meter}$. The interaction strength $g_s$ is recalibrated in this new geometry, following the slightly different confinement seen by the atoms. We then imprint a weak localized optical potential $U_\mathrm{ext}(x) = U_0 f(x)$ on both spin components characterized, in physical units, by $\sigma=\SI{2.1\pm0.1}{\micro\metre}$, $x_0 \gtrsim 0$, and $U_0 = \SI{-9\pm1}{\nano\kelvin}\times k_B$.  In the low-depletion limit, this depth corresponds to $\epsilon = (1-g_{1,2}/g)\, U_0/ (n_0 g_s) \approx -0.013\, U_0/ (n_0 g_s) \approx 0.1$. Owing to the minus sign in the proportionality coefficient between $\epsilon$ and $U_0$, we use $U_0 < 0$ in order to imprint an effective repulsive potential, which is more favorable for the fission of the multi-soliton\,\cite{Marchukov2019}. The potential amplitude is ramped linearly from zero to its final value over $\SI{100}{\milli\second}$. As a result, the initial two-soliton splits into two spatially separated wavepackets, as shown in Fig.~\ref{fig:splitting_exp}(a-b). We illustrate the time evolution of their fission in Fig.~\ref{fig:splitting_exp}(c), and extract the widths and atom numbers of each wavepacket, denoted $N_1$ and $N_2$. Fig.~\ref{fig:splitting_exp}(d) represents the fitted widths after the separation, which remain approximately constant, and in agreement with the expected magnetic solitons profiles.

In the NLSE limit, the soliton mass is proportional to its amplitude $\kappa = 2\, \mathrm{Im} (\lambda_j)$, so that the atom-number ratio directly reflects the ratio of scattering eigenvalues. It yields $N_2/N_1=\mathrm{Im}(\lambda_2/\lambda_1)=3$ for a NLSE two-soliton, and does not depend on the value of $\kappa$. This simple relation no longer holds for magnetic multi-solitons, for which the mass scales nonlinearly with the eigenvalues $\mathrm{Im} (\lambda_j)$. Using the expression for the mass of a single magnetic soliton [see Eq.~\eqref{eq:mag_soliton_N}], the splitting ratio instead reads:
\begin{equation}
    \frac{N_2}{N_1} = \frac{\tanh^{-1}(3 \kappa)}{\tanh^{-1}(\kappa)}\,,
    \label{eq:ratio}
\end{equation}
As a consequence, varying $\kappa$ leads to a continuous change in the splitting ratio, which we investigate in Table~\ref{tab:splitting_exp}. The explored values of $\kappa$ leads to ratios between $\approx\,$3 (the NLSE limit) and $\approx\,$5, in good agreement with Eq.~\eqref{eq:ratio}, which predicts a divergence of the ratio for $\kappa \to \kappa_c(2)=1/3$. This experiment thus realizes an analog IST of magnetic multi-soliton states, revealing quantitatively their constituents solitons.
\begin{table}
\begin{center}
\renewcommand{\arraystretch}{1.4}
\begin{tabular}{ >{\centering\arraybackslash}p{1.2cm}  >{\centering\arraybackslash}p{2.4cm} >{\centering\arraybackslash}p{2.4cm}} 
 \Xhline{1.2pt}
$ \kappa$& $N_2/N_1$ (exp.) & $N_2/N_1$ (theo.)\\ [0.1ex]
 \hline
 0 & -- & 3 \\ 
 0.20 & 3.2(2) & 3.42  \\
 0.25 & 3.6(2) & 3.79  \\
 0.31 &5.4(2) & 5.23  \\
 1/3 &-- &$\infty$  \\ 
 \Xhline{1.2pt}
\end{tabular}
\end{center}
\caption{\textbf{Splitting ratio.} Measured values of the splitting ratio $N_2/N_1$ for different  $\kappa$. The right column is the prediction of Eq.\,\eqref{eq:ratio}. Experimental values of $\kappa$ have a common uncertainty of $\pm 0.01$.}
\label{tab:splitting_exp}
\end{table}
%

\section{Summary and outlook}

In this work, we have demonstrated the deterministic experimental realization of magnetic multi-soliton states in a uniform quasi-one-dimensional immiscible two-component Bose gas. Exploiting the gauge equivalence between the LLE and the attractive NLSE, we constructed magnetic multi-solitons directly from well-known NLSE solutions and observed their characteristic breathing dynamics. 

By introducing a controlled perturbation, we were able to break integrability in a tunable manner and induce the fission of magnetic two-solitons into their fundamental constituents. This process reveals the composite structure of multi-soliton states and constitutes an experimental analog of the IST, in which individual solitons are identified. Our results highlight the degree of control offered by two-component Bose gases for exploring near-integrable nonlinear dynamics.

Beyond isolated multi-soliton states, our platform naturally opens perspectives toward the study of soliton gases, where a large number of interacting solitons form a statistical ensemble with emergent collective properties. Such regimes have recently attracted renewed interest in hydrodynamics and nonlinear optics\,\cite{Redor2019, Suret2024}, and remain largely unexplored in quantum gases. More broadly, the ability to engineer and manipulate integrable spin mixtures in a uniform 1D setting paves the way for investigations of integrable turbulence, extreme events, and the controlled generation of rogue-wave-like structures in ultracold gases. \\

\emph{Acknowledgments.} We acknowledge the support by ERC (Grant Agreement No 863880) and by ANR (ANR-23-PETQ-2700002). Y.L acknowledges funding by the LabEx ENS-ICFP: ANR-10-LABX-0010/ANR-10-IDEX-0001-02 PSL. We thank Thibault Congy and Nicolas Pavloff for fruitful discussions. We thank Raphael Lopes, Alexei Ourjoumtsev and Brice Bakkali-Hassani for their careful reading of the manuscript.
\newpage

\appendix

\section{IST details for NLSE}
\label{app:IST}

The Jost solutions $\Phi$ and $\Psi$ of the spectral problem Eq.~\eqref{eq:IST_eigen} are defined by their asymptotic boundary conditions:
\begin{subequations}
    \begin{alignat}{2}
        \Phi(x, t; \lambda) = \mqty(\phi_a \\ \phi_b) &\sim \mqty( 1 \\ 0) \e^{-\mi \lambda x -2 \mi \lambda^2 t} &&\qq{as} x \to -\infty\,,\\
        \Psi(x, t; \lambda) = \mqty(\psi_a \\ \psi_b) &\sim \mqty( 0 \\ 1) \e^{+\mi \lambda x +2 \mi \lambda^2 t} &&\qq{as} x \to +\infty\,.
    \end{alignat}
    \label{eq:IST_Jost}
\end{subequations}\unskip\ignorespaces
The quantities $\bar{\Phi}(x, t; \lambda^*) = \mqty(\phi_b^* & -\phi_a^*)^\intercal$ also satisfy Eq.~\eqref{eq:IST_eigen}, with the eigenvalue $\lambda^*$. The Jost solutions form two complete bases that are related by scattering coefficients\,\cite{Ablowitz1974}:
\begin{equation}
	(\Phi, \bar{\Phi}) = (\bar{\Psi}, \Psi)\, S \qq{and} S(\lambda) = \mqty(a & \hphantom{-}b^* \\ b & - a^*)\,,
	\label{eq:scattering_mat}
\end{equation}
where we have set $\lambda \in \mathbb{R}$, since part of the scattering data $a(\lambda), b(\lambda)$ is rigorously defined only on the real axis, and introduced the scattering matrix $S$. 
The field $u$ can be interpreted as an effective \enquote{potential} on which the plane-wave solutions $\Phi$ and $\Psi$ are partially transmitted and reflected, this information being encoding in the scattering data. 

More formally, the function $a$ is well-defined for $\lambda \in \mathbb{C}$ with $\mathrm{Im}(\lambda)>0$\,\cite{Zakharov1972_russ}. Its zeros $\lambda_j$, assumed to be simple, characterize localized Jost solutions, for which $\Phi(\lambda_j) = c_j \Psi(\lambda_j)$. The quantities $\lambda_j, c_j, b(\lambda)/a(\lambda)$ constitute the scattering data. Their time evolution is trivial, and can be deduced by combining Eq.~\eqref{eq:IST_Jost} and Eq.~\eqref{eq:scattering_mat}; $\lambda_j$ is time-independent, while $c_j(t) = \exp(4\mi \lambda_j^2 t)\, c_j(0)$ and $b(\lambda, t)/a(\lambda, t)= \exp(4\mi \lambda^2 t)\, b(\lambda, 0)/a(\lambda, 0)$.

Inverting the scattering data requires a Gelfand-Levitan-Marchenko integral equation:
\begin{equation}
    0 = \bar{K}(x,y) + F(x+y) \,\mqty(0 \\1)  + \int_x^{\infty} F(s+y) K(x, s)  \dd{s}\,,
    \label{eq:IST_Marchenko}
\end{equation}
where $K = \mqty(K_1 & K_2)^\intercal$ and $\bar{K} = \mqty(K_2^* & -K_1^*)^\intercal$, and we have defined:
\begin{equation}
    F(x) = \frac{1}{2\pi} \int_{-\infty}^{\infty} \frac{b(\lambda)}{a(\lambda)} \e^{\mi \lambda x} \dd{\lambda} - \mi \sum_{j=1}^{n} \tilde{c}_j \e^{\mi \lambda_j x} \,,
    \label{eq:IST_F}
\end{equation}
where $\tilde{c}_j = {c_j}/{a'(\lambda_j)}$. We have used $a'(\lambda_j)$, the derivative of $a(\lambda)$, evaluated at $\lambda_j$. Then, one obtains $u(x) = -2 K_1(x,x)$. 

When no radiative component is present, \textit{i.e.} $b(\lambda)=0$, a general formula for multi-solitons can be obtained (see \textit{e.g.} Ref.~\cite{Yang2010}):
\begin{equation}
    u(x,t) = -2\mi \sum_{j,k=1}^n \frac{1}{c_j(0)} \e^{\theta_j - \theta_k^*} \left(A^{-1}\right)_{j,k}\,,
    \label{eq:IST_multi_sol}
\end{equation}
where
\begin{equation}
	\left(A \right)_{j,k} = \frac{1}{\lambda_j^* - \lambda_k} \left[ \e^{-(\theta_k + \theta_j^*)} + \frac{1}{c_j^*(0) c_k(0)} \e^{\theta_k + \theta_j^*} \right]\,,
\end{equation}
and we introduced $\theta_j = -\mi \lambda_j x - 2\mi \lambda_j^2 t$.

\section{Coupled NLSEs to LLE}
\label{app:LLE}

In the Manakov regime, where $\abs{g_s} \ll g$, density and spin degrees of freedom decouple. As a result, the total density can be taken as constant, allowing the parametrization of Eq.~\eqref{eq:psi_param}. Imposing a vanishing total current then yields the constraint\,\cite{Rabec2025}:
\begin{equation}
    \chi_x = -\cos(\theta)\, \varphi_x\,,
    \label{eq:Phi_param}
\end{equation}
which eliminates the global phase $\chi$ in favor of the angular variables $\theta$ and $\varphi$.

Under these assumptions, the coupled NLSEs reduce to the equations of motion:
\begin{subequations}
    \begin{align}
        {\pdv{}{t}} \left[\cos(\theta)\right] &= -\pdv{}{x} \left[ \sin^2(\theta) \pdv{\varphi}{x} \right] \\
        \sin(\theta) \pdv{\varphi}{t} &= -\pdv[2]{\theta}{x} + \sin(\theta) \cos(\theta) \left[1 + \left( \pdv{\varphi}{x} \right)^2 \right]\,,
    \end{align}
    \label{eq:GPE_polar}
\end{subequations}\unskip\ignorespaces
which correspond to the Landau-Lifshitz equation for a one-dimensional easy-axis ferromagnet\,\cite{Kosevich1990, Rabec2025}.

Physical units can be restored in Eq.~\eqref{eq:coupled_NLSE} by introducing the rescaled variables $\tilde{x} = x \,\xi_s$, $\tilde{t}= t \,\tau_s$, and $\tilde{u}_i = u_i \sqrt{n_0}$, where $\tau_s=\hbar/(g_s n_0)$.

\section{Magnetic soliton}
\label{app:Mag_soliton}

Closed-form expressions for easy-axis magnetic solitons of the Landau-Lifshitz equation are available in the literature (see, e.g., Refs.~\cite{Borovik1988, Kosevich1990, Rabec2025}). They can be written as a function of two parameters: $v$, denoting the soliton velocity, and $\Omega$, an internal precession frequency. The magnetic soliton corresponding to the $n=1$ NLSE soliton given by Eq.~\eqref{eq:IST_bright_sol} can be obtained, up to position offset, using Eq.~\eqref{eq:diff_eq_theta_phi}:
\begin{equation}
    \left\{
    \begin{aligned}
        \cos[\theta(x, t)] & = 1  - \frac{4 \kappa^2}{2 + \Omega + \sqrt{\Omega^2 + v^2} \cosh(2 \kappa \tilde{x})} \\
        \Omega t - \varphi(x, t) & = \frac{1}{2} v \tilde{x} + \arctan\left[\frac{2 \Omega - v^2 + 2\sqrt{\Omega^2 + v^2}}{2 \kappa v} \, \tanh(\kappa \tilde{x}) \right] \,,
    \end{aligned}
    \right.
    \label{eq:mag_soliton}
\end{equation}
where $\kappa = \sqrt{1 + \Omega - v^2/4}$ and $\tilde{x} = x-vt$. Solutions exist only when the condition $ \Omega - v^2/4 > -1$ is satisfied. 

These solutions are characterized by several conserved quantities. Of particular relevance here is the soliton \emph{mass} $N$, which can be written as:
\begin{equation}
	\tanh(N/2) = \frac{2+\Omega - \sqrt{\Omega^2+v^2}}{2\kappa}\,.
	\label{eq:mag_soliton_N}
\end{equation}
In this work, we restrict to stationary solutions ($v=0$) in the regime $\Omega<0$, which directly yields Eq.~\eqref{eq:ratio}. The corresponding dimensional atom number in the minority component is obtained as $N_\mathrm{at} = N \xi_s n_0$.

\begin{figure*}[t!!!]
    \begin{center}
	   \includegraphics[]{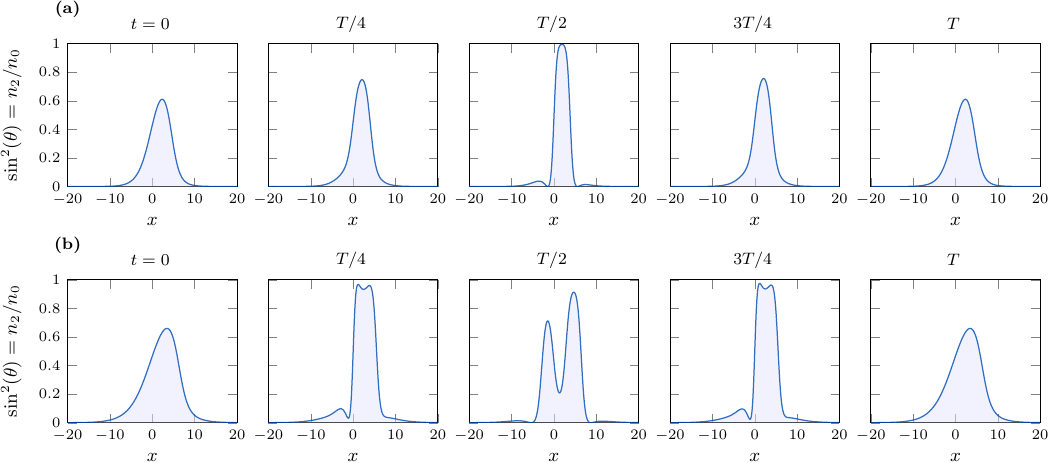} 
    \end{center}
\caption{\textbf{Numerical time evolution of magnetic $n$-solitons.} We represent the normalized density of the minority component, $n_2/n_0$ obtained by combining Eq.~\eqref{eq:N_sol_NLSE} and the numerical resolution of the differential Eq.~\eqref{eq:diff_eq_theta_phi}. Equivalently, one could also determine the full time of evolution of the system by solving LLE. We show in \textbf{(a)} some remarkable times of the evolution of a two-soliton, computed with $\kappa=0.31$. \textbf{(b)} Same as in \textbf{(a)}, for a three-soliton and $\kappa=0.195$. Each period $T$ can be obtained from Eq.~\eqref{eq:sol_freq}.}
    \label{fig:numerics}
\end{figure*}

\section{Gauge equivalence}
\label{app:Gauge}

We note the Lax pair $(\hat{L}, \hat{A})$ of the LLE. Both operators are polynomial in the spectral parameter $\lambda$. We denote by $\hat{L}^{(i)}$ (resp. $\hat{A}^{(i)}$) the coefficient multiplying $\lambda^i$ in $\hat{L}$ (resp. $\hat{A}$). We write similarly the Lax operators of the NLSE, denoted with primes. Expanding Eq.~\eqref{eq:gauge_eq} order by order in $\lambda$ yields five equations:
\begin{subequations}
	\begin{align}
	    \hat{L}^{(0)} &= \hat{G} \hat{L}'^{(0)} \hat{G}^{-1} + \hat{G}_x \hat{G}^{-1} \label{eq:gauge_eq1}\,,\\
	    \mi \hat{M} &= \mi \hat{G} \hat{\sigma}^{(z)} \hat{G}^{-1} \label{eq:gauge_eq2}\,,\\
	    \hat{A}^{(0)} &= \hat{G} \hat{A}'^{(0)} \hat{G}^{-1} + \hat{G}_t \hat{G}^{-1} \label{eq:gauge_eq3}\,,\\
	    \hat{M} \hat{M}_x + 2\hat{L}^{(0)} &= 2\hat{G} \hat{L}'^{(0)} \hat{G}^{-1} \label{eq:gauge_eq4}\,,\\
	    2\mi \hat{M} &= 2\mi \hat{G} \hat{\sigma}^{(z)} \hat{G}^{-1}\,.\label{eq:gauge_eq5}
	\end{align}
\end{subequations}\unskip\ignorespaces
Combining Eq.~\eqref{eq:gauge_eq1} and \eqref{eq:gauge_eq2}, one finds:
\begin{equation}
    \hat{M} \hat{M}_x = -2 \hat{G}_x \hat{G}^{-1}\,,
\end{equation}
using the identities $\hat{M} \hat{L}^{(0)} \hat{M} = - \hat{L}^{(0)}$ and $\hat{\sigma}^{(z)} \hat{L}'^{(0)} \hat{\sigma}^{(z)} = - \hat{L}'^{(0)}$. Then, Eq.~\eqref{eq:gauge_eq4} follows automatically. Thus, we aim to construct $\hat{G}$ such that Eqs.~\eqref{eq:gauge_eq1} and \eqref{eq:gauge_eq2} hold for an initial condition $u(x,0)$, which directly yields the corresponding magnetization $\bs{M}(x,0)$. These conditions can be combined into:
\begin{equation}
    \hat{G}^{-1} \hat{G}_x = -\hat{L}'^{(0)} + \frac{1}{4} \left[\hat{G}^{-1} \hat{\sigma}^{(z)} \hat{G}, \hat{\sigma}^{(z)} \right]\,.
    \label{eq:diff_eq_g}
\end{equation}
Since $\hat{G} \in \mathrm{SU}(2)$\,\cite{Kundu1984}, it can be parametrized by two functions $g_1$ and $g_2$ as:
\begin{equation}
    \hat{G} = \mqty(g_1^* & -g_2 \\ g_2^* & \hphantom{-}g_1)\,.
\end{equation}
Substituting this form into Eq.~\eqref{eq:diff_eq_g} and extracting the upper-right entry gives:
\begin{equation}
    u(x, t) = g_1 g_{2x} - g_2 g_{1x} + g_1 g_2\,.
\end{equation}
Next, we express $g_1$ and $g_2$ in terms of the polar angles of $\bs{M}$, using in particular $\hat{M} = \hat{G} \hat{\sigma}^{(z)} \hat{G}^{-1}$ and $\{\hat{G}^{-1}\hat{G}_x, \hat{\sigma}^{(z)} \} =0$, where $\{\cdot,\cdot\}$ is the anticommutator, and the boundary conditions. This results in:
\begin{equation}
	\begin{split}
    \hat{G}(x,t) = \mqty(\cos{\frac{\theta}{2}} \, \e^{-\mi \delta} & -\sin{\frac{\theta}{2}} \, \e^{\mi (\delta-\varphi)}\\ \sin{\frac{\theta}{2}} \, \e^{-\mi (\delta-\varphi)} & \cos{\frac{\theta}{2}} \, \e^{\mi \delta}) \\  \qq{with} \delta(x,t) = \int_{-\infty}^x \sin^2{\left(\frac{\theta}{2}\right)} \,\varphi_y \dd{y} \,.
    \label{eq:B_g_expl}
    \end{split}
\end{equation}
Thus, we obtain the differential equation:
\begin{equation}
    u(x,t) = \frac{1}{2} \left[\theta_x + \sin{\theta} \left(1-\mi \varphi_x  \right) \right] \exp(-\mi \int_{-\infty}^x \cos{\left(\theta\right)}\, \varphi_y \dd{y} ) \,.
    \label{eq:diff_eq_theta_phi}
\end{equation}
In the case of a real field $u$, it reduces to Eq.~\eqref{eq:diff_eq_theta} of the main text. As an example, we show in Fig.\,\ref{fig:numerics}, the time evolution of given magnetic two- and three-solitons.

\newpage 

\definecolor{violet3}{RGB}{70,0,120}
\hypersetup{urlcolor=violet3} 

\bibliography{Bib_solitonsLLE}

\end{document}